\documentclass[aps,onecolumn,preprintnumbers,amsmath,amssymb,nofootinbib,superscriptaddress,notitlepage,11pt]{revtex4-1}

\usepackage{inputenc}
\usepackage{txfonts}
\usepackage{epsfig}
\usepackage{slashed}
\usepackage{color}
\usepackage{overpic}
\usepackage{subfigure}
\usepackage{hyperref}

\begin{document}

\title{$D_{s0}(2590)$ as a dominant $c\bar{s}$ state with a small  $D^*K$ component}

\author{Jia-Ming Xie}
\affiliation{School of Physics, Beihang University, Beijing 102206, China}

\author{Ming-Zhu Liu}
\affiliation{School of Space and Environment, Beihang University, Beijing 102206, China}
\affiliation{School of Physics, Beihang University, Beijing 102206, China}

\author{Li-Sheng Geng}\email{lisheng.geng@buaa.edu.cn}
\affiliation{School of Physics, Beihang University, Beijing 102206, China}
\affiliation{
Beijing Key Laboratory of Advanced Nuclear Materials and Physics,
Beihang University, Beijing 102206, China}
\affiliation{School of Physics and Microelectronics, Zhengzhou University, Zhengzhou, Henan 450001, China}

\date{\today}
\begin{abstract}

    The recently discovered $D_{s0}(2590)$ state by the LHCb collaboration was regarded as the first excited state of $^1S_{0}$ charmed-strange meson. Its mass is, however, lower than  the Godfrey-Isgur quark model prediction by about 80 MeV. In this work, we take into account the $D^{\ast}K$ contribution to the bare $c\bar{s}$ state, and show that the coupled-channel interaction induces an 88 MeV shift with respect to the conventional quark model $c\bar{s}$ state, which is much closer to the experimental mass. Our study shows that  in addition to $S$-wave, $P$-wave coupled-channel interactions  also play a role for hadrons located close to two-hadron thresholds. We  further scrutinize the unquenched quark model results  with a model independent approach. It is shown  that  the two-body $D^*K$ decay width is proportional to the weight of the $D^*K$ component. To saturate the experimental total decay width with the $D^*K$ partial decay width we need a weight of about 60\% while to reproduce the unquenched quark model result a weight of about 5\% is needed. Therefore, we  encourage  future experimental studies on the two-body $D^*K$ partial decay   of $D_{s0}(2590)$.

\end{abstract}


\maketitle

\section{Introduction}

 The quark model, which has been rather successful in describing the properties of ground-state hadrons, was challenged in 2003 by the discovery of $X(3872)$~\cite{Belle:2003nnu,CDF:2003cab,D0:2004zmu,BaBar:2004oro} and $D_{s0}^*(2317)$~\cite{BaBar:2003oey,CLEO:2003ggt,Belle:2003kup}. Since then, a large number of states that cannot be easily explained by the quark model were observed, most of which lie close to the mass threshold of a pair of hadrons~\cite{Chen:2016qju,Hosaka:2016ypm,Lebed:2016hpi,Guo:2017jvc,Olsen:2017bmm,Ali:2017jda,Brambilla:2019esw,Liu:2019zoy}. Thus it is natural to expect that these states contain large hadronic molecular components. However, for certain systems,  a more refined picture considering hadronic molecular components, conventional $q\bar{q}$ or $qqq$ components, and compact multiquark components is needed. In the unquenched quark model, coupled-channel effects are taken into account on top of the conventional $q\bar{q}$ or $qqq$ configuration--so-called bare state. In the unquenched quark model, the interaction between the bare state and the coupled channels can be described by the $^3P_{0}$ mechanism~\cite{Tan:2019qwe,Luo:2019qkm,Luo:2021dvj}. 
 The $^3P_{0}$ model, also referred to as the vacuum quark-pair creation model, was originally proposed by Micu~\cite{Micu:1968mk}, and then  further developed by A. Le Yaouanc. et al.~\cite{LeYaouanc:1972vsx,LeYaouanc:1973ldf}. It has been  widely used to investigate the Okubo-Zweig-Iizuka~(OZI)-allowed two-body strong decays of conventional hadrons~\cite{Barnes:2005pb,Godfrey:2015dia,Godfrey:2015dva}. The quark-pair creation model also provides an approach to construct the interaction between the initial state and the subsequent two-body strong decay channel. With the transition amplitude provided by the $^3P_{0}$ model and conventional quark model, an unquenched quark model can be constructed, which takes into account the coupled-channel effects. In particular, the unquenched quark model can give a quantitative estimate of the ratio between the hadronic molecule components and  the bare state. For instance, in the unquenched quark model, $X(3872)$ was well described as a mixture  of about 70\%  $\chi_{c1}(2P)$ and 30\% $D^{\ast}\bar{D}$~\cite{Tan:2019qwe}. 
With a similar approach, Luo, et al. solved the low mass puzzle of $\Lambda_{c}(2940)$ between the naive quark model prediction and the experimental mass by considering the contribution of the $D^{\ast}N$ channel, which strongly couples to the $\Lambda_{c}(2P,3/2^{-})$ bare state~\cite{Luo:2019qkm}.
In Ref.~\cite{Luo:2021dvj}, the authors considered the $DK$ contribution to $D_{s0}^{\ast}(2317)$, leading to a 77 MeV mass shift with respect to the conventional quark model prediction. Very recently, Yang et al. used the Hamiltonian effective field theory to study the mass spectrum of positive parity $D_{s}$ resonant states considering both the $P$-wave $c\bar{s}$ core and the $DK$/$D^{\ast}K$ coupled channels and found that the $D_{s0}^*(2317)$ state contains about 60\% $DK$ and 40\% $c\bar{s}$~\cite{Yang:2021tvc}.

A common feature of the above mentioned exotic states is that the orbital angular momentum between the two quark components is $P$-wave. The $P$-wave excitation is equal to the creation of a pair of quark and antiquark  in vacuum, which can form two hadrons with its original quark components through the quark rearrangement mechanism. If the mass threshold of these two hadrons is close to the mass of the bare state, it would couple to the bare state and lower its mass, leading to a smaller mass for the physical state in agreement with the experimental measurement.

Recently, the LHCb Collaboration observed a new excited $D_s^+$ state in the  $D^+K^+\pi^-$  mass distribution of the  $B^0 \rightarrow D^-D^+K^+\pi^-$ decay using a data sample corresponding to an integrated luminosity of 5.4 $\rm{fb}^{-1}$  at a centre-of-mass energy of 13 TeV~\cite{LHCb:2020gnv}. Its  mass,  width, and spin-parity are determined to be $m_R = 2591\pm6\pm7$ MeV, $\Gamma_R = 89 \pm 16\pm 12$ MeV, and $J^P=0^-$, respectively.  This state was considered to be a  candidate for the $D_s(2{}^1S_0)$ state, the radial excitation of the ground state $D_s$ meson. Nonetheless, the relativized quark model predicted that such a state should have a mass of 2673 MeV~\cite{Godfrey:2015dva}, which is larger than the experimental mass by 82 MeV.
One should note that there exists only one channel, $D^{\ast}K$, that could couple to the $D_s(2{}^1S_0)$ state in terms of the quark-pair creation mechanism.  In this work, we employ the unquenched quark model to investigate whether by considering the $D^{\ast}K$ contribution one can reconcile the experimental mass of $D_{s0}(2590)$ with that of the theoretical $D_s(2{}^1S_0)$ state. It should be noted that different from the cases of $X(3872)$, $D_{s0}^{\ast}(2317)$, and $\Lambda_c(2940)$, the $D^{\ast}K$ channel couples to the $c\bar{s}$ bare state via $P$-wave. 

In Ref.~\cite{Aceti:2014ala}, it was shown that the $\Delta(1232)$ state contains a substantial $\pi N$ component, about 60\%, using an extension of the Weinberg compositeness condition on partial waves of $\ell=1$. We will adopt the same approach to estimate the weight of the $D^*K$ component in the $D_{s0}(2590)$ state and compare it with the results of the unquenched quark model.

The paper is structured as follows. In Sec.~\ref{sec1} we present the details of the unquenched quark model as well as a brief description of the model-independent approach. In Sec.~\ref{sec3} we study the mass shift of the bare $D_s(2{}^1S_0)$  state induced by the coupling to $D^*K$ in the unquenched quark model and calculate its partial decay width into $D^{\ast}K$. Furthermore, we adopt the model independent approach to evaluate the weight  of $D^*K$ in the physical state $D_{s0}(2590)$.
Finally we present the conclusions in Sec.~\ref{sec4}.

\section{Theoretical formalism}
\label{sec1}

In this work, we adopt two different theoretical models to evaluate the relevance of the $D^*K$ coupled channel in the physical $D_{s0}(2590)$ state. The first is an unquenched quark model in the line of Ref.~\cite{Kalashnikova:2005ui} and the second is a model-independent approach in the line of Ref.~\cite{Aceti:2014ala}. In the following, we briefly describe these two methods. More details can be found in Refs.~\cite{Kalashnikova:2005ui,Aceti:2014ala}. 

\subsection{Unquenched quark model}
We first introduce the Hamiltonian in the framework of the unquenched quark model, which contains three terms  
\begin{equation}
    H_{D_{s0}(2590)}=H_{c\bar{s}}+H_{D^{\ast}K}+H_I, 
\end{equation}
where  $H_{c\bar{s}}$ comes from the conventional quark model~\cite{Godfrey:1985xj}, $H_{D^{\ast}K}$ denotes the interaction between $D^{\ast}$ and $K$, and $H_{I}$ stands for the interaction between the bare $ c\bar{s}$ state and the $D^{\ast}K$ channel. With the above Hamiltonian the wave function of the physical $D_{s0}(2590)$ state  can be written as  
\begin{equation}
    \Psi(D_{s0}(2590)) =c_{c\bar{s}}\Psi(c\bar{s})+\int d^3 p c_{D^{\ast}K}(\textbf{p})\Psi_{\textbf{p}}(D^{\ast}K),
    \label{eq6}
\end{equation}
which indicates that there are two Fock components, a $c\bar{s}$ core at quark level and a $D^{\ast}K$ component at hadron level. In the following, we specify each of the two terms. 
  The Hamiltonian $H_{c\bar{s}}$ is taken from the Godfrey-Isgur relativized potential quark model,
\begin{equation}
    H_{c\bar{s}}\Psi(c\bar{s})=M_0\Psi(c\bar{s}),
\end{equation}
where $M_{0}$ is the bare mass.  The term  $H_{D^{\ast}K}$ denotes the Hamiltonian of the  $D^{\ast}K$ system. As we only consider the kinetic energy but neglect the interaction between $D^{\ast}$ and $K$~\footnote{According to chiral perturbation theory, the $S$-wave $D^{\ast}K$ interaction is strong, but the $P$-wave interaction is weak~\cite{Altenbuchinger:2013vwa}.}, it can be written as
\begin{equation}
    H_{D^{\ast}K}\Psi_{\textbf{p}}(D^{\ast}K)=\left(\sqrt{m_{D^{\ast}}^2+|\textbf{p}|^2}+\sqrt{m_{K}^2+|\textbf{p}|^2}\right)\Psi_{\textbf{p}}(D^{\ast}K),
\end{equation}
where  $\textbf{p}$ represents the 3-momentum of $D^{\ast}$  or $K$ meson  in the center-of-mass frame of $D^{\ast}K$.

In the quark-pair creation model~(of which the details are relegated to Appendix~\ref{Appendix}), the  transition operator of $c\bar{s}(2^1S_0)\rightarrow D^{\ast}K$ is written as 
\begin{equation}
    T=-3\gamma \omega_0 \phi_0 \sum_{m=-1,0,1} C_{11}(00; m -m) \chi_{1,-m} \int d^3 p_{q} \int d^3 p_{\bar{q}} \delta^{(3)}(\textbf{p}_q+\textbf{p}_{\bar{q}}) \mathcal{Y}^m_1\left(\frac{\textbf{p}_q-\textbf{p}_{\bar{q}}}{2}\right) b_q^{\dagger}(\textbf{p}_q) d_{\bar{q}}^{\dagger}(\textbf{p}_{\bar{q}}),
    \label{eq5}
\end{equation}
where $\omega_0$, $\phi_0$, and $\chi_{1,-m}$ are the SU(3)-color singlet, SU(3)-flavor singlet and  spin triplet wave function, and $\mathcal{Y}_\ell^m(\textbf{p})=|\textbf{p}|^\ell Y_{\ell}^{m}(\textbf{p})$ is the solid harmonics.  The single dimensionless free parameter $\gamma$ describes the strength of the creation of the $q\bar{q}$ pair.  The Clebsch-Gordan coefficients $C_{11}(00; m -m)$ denote the coupling of the spin and the orbital angular momenta of the $q\bar{q}$ pair into total spin $0$.
 The delta function  $\delta^{(3)}(\textbf{p}_q+\textbf{p}_{\bar{q}})$  constrains the momentum of the $q\bar{q}$ pair, in agreement with the quark-pair creation in vacuum. Accordingly  $b_{q}^{\dagger}$ and $d_{\bar{q}}^{\dagger}$ are quark and antiquark creation operators, respectively.
 Thus the Hamiltonian $H_I$ between $D^{\ast}K$ and the $c\bar{s}$ core can be expressed as
\begin{equation}
    H_I=T+T^{\dagger}.
\end{equation}
The  resulting eigenvalue equation has the following form 
 \begin{equation}
\left( \begin{array}{cc} \Big(\Psi(c\bar{s}),H_{c\bar{s}}\Psi(c\bar{s})\Big) & \int d^3 p \left(\Psi(c\bar{s}),T^{\dagger}\Psi_{\textbf{p}}(D^{\ast}K)\right) \\ \left(\Psi_{\textbf{p}}(D^{\ast}K),T\Psi(c\bar{s})\right) & \left(\sqrt{m^2_{D^{\ast}}+|\textbf{p}|^2}+\sqrt{m^2_{K}+|\textbf{p}|^2}\right) \end{array} \right) \left( \begin{array}{cc} c_{c\bar{s}} \\ c_{D^{\ast}K}(\textbf{p}) \end{array} \right) \\ =M\left( \begin{array}{cc} c_{c\bar{s}} \\ c_{D^{\ast}K}(\textbf{p}) \end{array} \right).
\end{equation}
where we assume that the two Fock states, $\Psi(c\bar{s})$ and $\Psi_{\textbf{p}}(D^{\ast}K)$, are orthogonal to each other and properly normalized respectively.  The non-diagonal term is  the 
transition amplitude
\begin{equation}
\label{amplitude}
    \mathcal{M}_{c\bar{s}(2^1S_0)\rightarrow D^{\ast}K(\textbf{p})}(\textbf{p})\equiv \left(\Psi_{\textbf{p}}(D^*K),T\Psi(c\bar{s})\right). 
\end{equation}
The above matrix equation can be simplified to two algebraic equations 
\begin{equation}
    \begin{aligned}
     M_0 c_{c\bar{s}}+\int d^3 p \mathcal{M}_{c\bar{s}(2^1S_0)\rightarrow D^{\ast}K(\textbf{p})}^{\ast}(\textbf{p}) c_{D^{\ast}K}(\textbf{p})&=M c_{c\bar{s}}, \\
    \mathcal{M}_{c\bar{s}(2^1S_0)\rightarrow D^{\ast}K(\textbf{p})}(\textbf{p}) c_{c\bar{s}}+\left(\sqrt{m^2_{D^{\ast}}+|\textbf{p}|^2}+\sqrt{m^2_{K}+|\textbf{p}|^2}\right) c_{D^{\ast}K}(\textbf{p})&=M c_{D^{\ast}K}(\textbf{p}).
    \end{aligned}
    \label{eq1}
\end{equation}
From Eq.~(\ref{eq1}) we can derive the following relation
\begin{equation}
    M-M_0-\Delta M(M)=0,
    \label{eq2}
\end{equation}
where the mass shift $\Delta M(M)$ is defined as 
\begin{equation}
    \label{eq3}
    \Delta M(M)\equiv \textrm{Re} \int d^3 p \frac{\left|\mathcal{M}_{c\bar{s}(2^1S_0)\rightarrow D^{\ast}K(\textbf{p})}(\textbf{p})\right|^2}{M-\sqrt{m^2_{D^{\ast}}+|\textbf{p}|^2}-\sqrt{m^2_{K}+|\textbf{p}|^2}+i\epsilon}.
\end{equation}
From Eq.~(\ref{eq2}) and Eq.~(\ref{eq3})  the physical mass $M$ and the mass shift $\Delta M$ can be determined simultaneously. The coupled-channel correction to the bare state is the mass shift $\Delta M$. 

In order to estimate the $c\bar{s}$ core contribution to the physical state $D_{s0}(2590)$,  we need to calculate the $Z$-factor, i.e., the field renormalization constant~\cite{Weinberg:1965zz}, defined as 
\begin{equation}
    Z\equiv\left|\Big(\Psi(c\bar{s}),\Psi(D_{s0}(2590))\Big)\right|^2,
    \label{eq9}
\end{equation}
where $\Psi(c\bar{s})$
represents the genuine $c\bar{s}$ component of the state. If we ignore the narrow decay width $\Gamma$ compared with its mass $m_R = 2591 \pm 6 \pm 7$ MeV of the $D_{s0}(2590)$ state, we can write down the normalization condition
\begin{equation}
    1=\Big( \Psi(D_{s0}(2590)), \Psi(D_{s0}(2590)) \Big).
    \label{eq10}
\end{equation}
With two Fock components in $D_{s0}(2590)$ (see Eq.~(\ref{eq6})) the normalization condition becomes 
\begin{equation}
    1=\left| c_{c\bar{s}} \right|^2 +\int d^3 p \left| c_{D^{\ast}K}(\mathbf{p}) \right|^2.
    \label{eq7}
\end{equation}
From Eq.~(\ref{eq1}), we can obtain the relation between the wave function of the $c\bar{s}$ core and that of $D^{\ast}K(\mathbf{p})$
\begin{equation}
    c_{D^{\ast}K}(\mathbf{p})=\frac{\mathcal{M}_{c\bar{s}(2^1S_0)\rightarrow D^{\ast}K(\textbf{p})}(\textbf{p})}{M-\sqrt{m^2_{D^{\ast}}+|\textbf{p}|^2}-\sqrt{m^2_{K}+|\textbf{p}|^2}+i\epsilon}c_{c\bar{s}}.
\end{equation}
With this,  Eq.~(\ref{eq7}) becomes
\begin{equation}
    1=\left| c_{c\bar{s}} \right|^2 \left( 1+\mathrm{Re} \int d^3 p \frac{\left| \mathcal{M}_{c\bar{s}(2^1S_0)\rightarrow D^{\ast}K(\textbf{p})}(\textbf{p}) \right|^2 }{\left( M-\sqrt{m^2_{D^{\ast}}+|\textbf{p}|^2}-\sqrt{m^2_{K}+|\textbf{p}|^2}\right)^2+i\epsilon}\right).
\end{equation}
Finally we obtain the $Z$-factor
\begin{equation}
    Z=\left| c_{c\bar{s}} \right|^2=\left( 1+\mathrm{Re} \int d^3 p \frac{\left| \mathcal{M}_{c\bar{s}(2^1S_0)\rightarrow D^{\ast}K(\textbf{p})}(\textbf{p}) \right|^2 }{\left( M-\sqrt{m^2_{D^{\ast}}+|\textbf{p}|^2}-\sqrt{m^2_{K}+|\textbf{p}|^2}\right)^2+i\epsilon}\right)^{-1}.
    \label{eq8}
\end{equation}
A few remarks are in order. First, the integral in the above equation is only well defined for   $M<m_{D^*}+m_K$. However, in the original $^3P_0$ model, the physical state is supposed to decay strongly into the two-body final state. Therefore, there seems to be an internal inconsistency  in the unquenched quark model specified above. The solution is quite straightforward. One should replace the real $M$ with its complex counterpart $M+i\Gamma/2$. Then Eq.~(\ref{eq8}) becomes
\begin{equation}
    Z=\left( 1+\mathrm{Re} \int d^3 p \frac{\left| \mathcal{M}_{c\bar{s}(2^1S_0)\rightarrow D^{\ast}K(\textbf{p})}(\textbf{p}) \right|^2 }{\left( M+i\Gamma/2-\sqrt{m^2_{D^{\ast}}+|\textbf{p}|^2}-\sqrt{m^2_{K}+|\textbf{p}|^2}\right)^2}\right)^{-1}.
\end{equation}
For self-consistency, one can also replace $M$ with $M+i\Gamma/2$ in the definition of the mass shift in Eq.~(\ref{eq3}), which would just cause tiny changes to $\Delta M$ since $\Gamma \ll M$ and then seems unnecessary. Note that $\Gamma$ here only represents the two-body strong decay width into $D^{\ast}K$ rather than the total experimental decay width $\Gamma_R = 89 \pm 16\pm 12$ MeV.

\subsection{Model independent approach}
One can also study the contributions of different Fock components to a physical state in a model independent way. In Refs.~\cite{Aceti:2014ala,Aceti:2012dd}, it was shown that the $\pi N$ component in the $\Delta(1232)$ state is substantial while the $\pi\pi$ component in the $\rho$ wave function is small.  In the following, we adopt such a method to  estimate the relative weights of the $c\bar{s}$ core and $D^{\ast}K$ coupled channel in  the physical $D_{s0}(2590)$. We first briefly introduce the essential ingredients of this approach.

The starting point is to parameterize the  $D^{\ast} K$ potential. Close to threshold, it has the following form
\begin{equation}
    v(s)=-\alpha\left(1+\beta \frac{s}{M_0^2-s}\right),
    \label{eq11}
\end{equation}
where $s$ is the square of the center-of-mass energy, and $M_{0}$ is the mass of the bare $c\bar{s}$ core. $\alpha$ and $\beta$ are two unknown  parameters that should  be determined by fitting to experimental data. In Refs.~\cite{Aceti:2014ala,Aceti:2012dd}, for the cases of the $\Delta(1232)$ and $\rho$, such unknown parameters are fixed by fitting to the $\pi N$ and $\pi \pi$ scattering data. 

The above potential can then be inserted into the  Lippmann-Schwinger equation to obtain the transition amplitudes
\begin{equation}
    t(s)=\frac{1}{v^{-1}(s)-G(s)},
\end{equation}
where $G(s)$ is the loop function of $D^{\ast}$ and $K$ mesons, which has the following form  to account for the $P$-wave nature of the $D^*K$ interaction
\begin{equation}
    G(s)=\int_{|\mathbf{p}| < \Lambda} \frac{d^3 p}{(2\pi)^3} \frac{|\mathbf{p}|^2}{s-(\omega_{D^{\ast}}+\omega_K)^2+i\epsilon} \left(\frac{\omega_{D^{\ast}}+\omega_K}{2\omega_{D^{\ast}}\omega_K}\right),
    \label{Gs}
\end{equation}
where $\omega_{D^{\ast}}=\sqrt{m_{D^{\ast}}^2+|\mathbf{p}|^2}$, $\omega_K=\sqrt{m_{K}^2+|\mathbf{p}|^2}$, and $\Lambda$ is the cutoff needed to regularize the integral.

Due to the fact that the physical $D_{s0}(2590)$ state is above the mass threshold of $D^{\ast}K$, we need the loop function in the second Riemann sheet, which is  defined as 
\begin{equation}
    G^{II}(s)=G^{I}(s)+i\frac{|\mathbf{p}|^3}{4\pi \sqrt{s}},
\end{equation}
where $G^{I}(s)$ is given in Eq.~(\ref{Gs}).

\section{Results and discussion}
\label{sec3}

For the mass $M_0$ of the bare $c\bar{s}$ state, we adopt the result of the Godfrey-Isgur relativized potential quark model~\cite{Godfrey:1985xj}
\begin{equation}
    M_0\left(c\bar{s}(2^1S_0)\right)=2673~\textrm{MeV},
\end{equation}
as well as the same set of parameters for the constituent quark masses:
\begin{equation}
   m_c=1628~\textrm{MeV},~ m_s=419~\textrm{MeV},~ m_{u/d}=220~\textrm{MeV}.
\end{equation} 
The masses of $D^*$ and $K$ are taken from the PDG~\cite{Zyla:2020zbs}:
\begin{equation}
   m_{D^{\ast}}=2008~\textrm{MeV},~m_{K}=495~\textrm{MeV}.
\end{equation} 
For the effective  simple harmonic oscillator parameter $\beta_{\rm{eff}}$ of the $c\bar{s}$ core and $D^{\ast}$, we choose~\cite{Godfrey:2015dia} 
\begin{equation}
   \beta_{\rm{eff}}\left(c\bar{s}(2^1S_0)\right)=0.475\ \textrm{GeV},\ \beta_{\rm{eff}}(D^{\ast})=0.516\ \textrm{GeV}.
\end{equation} 
Instead of taking $\beta_{\rm{eff}}$ to be 0.4 GeV for all light mesons as in Ref.~\cite{Godfrey:2015dia}, we choose the value determined in Ref.~\cite{Godfrey:1986wj} for $K$, 
\begin{equation}
   \beta_{\rm{eff}}(K)=0.710\ \textrm{GeV},
\end{equation} 
which obeys the uniform standards~\footnote{Stated below Eq.~(\ref{sho}).} for all involved mesons.
Finally, for the dimensionless vacuum $u\bar{u}/d\bar{d}$ quark-pair creation strength constant $\gamma$, we choose~\cite{Song:2015nia} 
\begin{equation}
    \gamma=8.7.
\end{equation}

\begin{figure}[h!]
    \centering
    \includegraphics[scale=0.5]{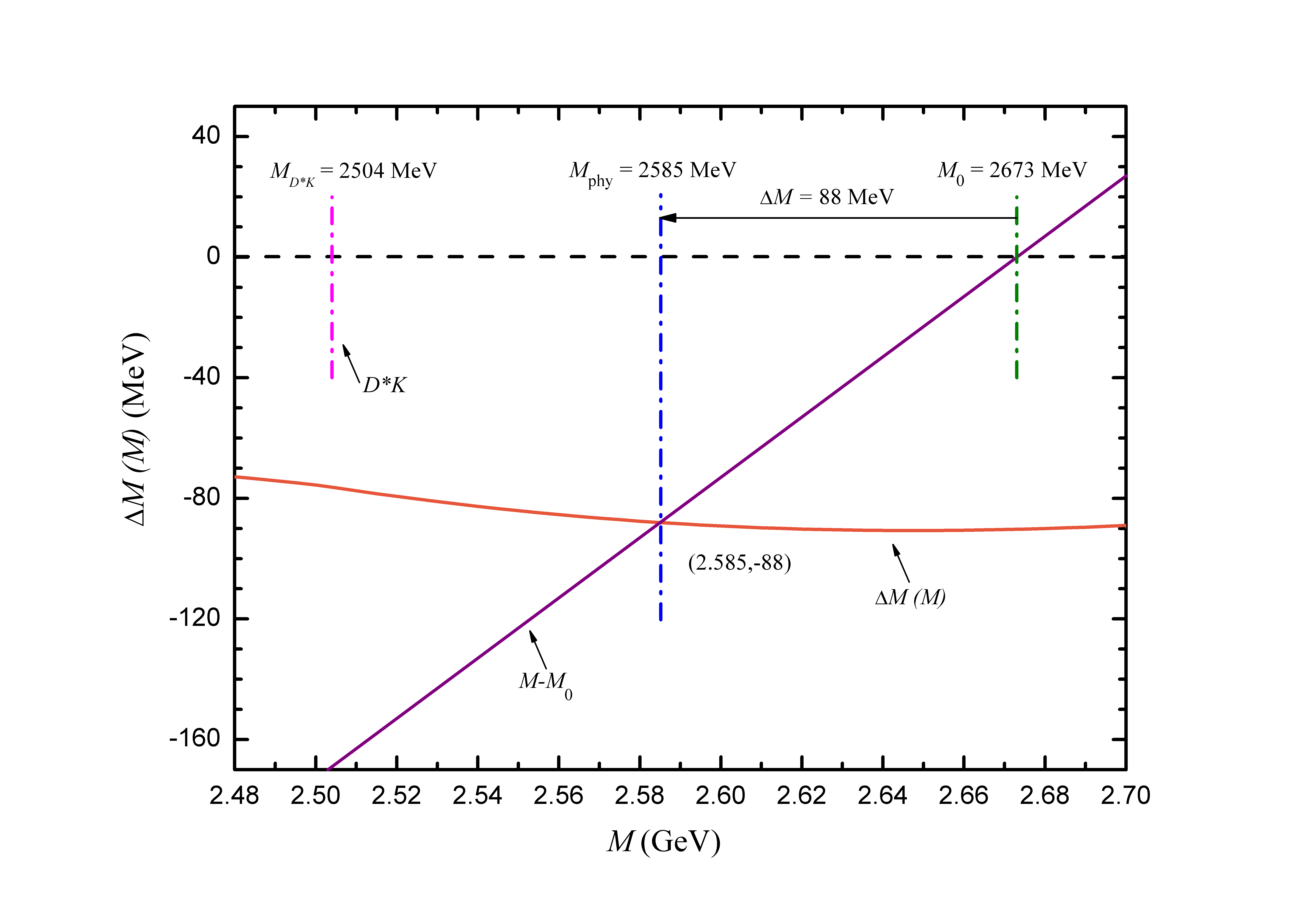}
    \caption{$\Delta M-M$ of $D_{s0}(2^1S_0)$, where the physical mass $M_{\rm{phy}}(D_{s0}(2^1S_0))$ is located at the intersection point of two solid lines.}
    \label{fig3}
\end{figure}

With the parameters specified above we can straightforwardly obtain the relation between the mass shift $\Delta M$ and the physical mass $M$, which is shown in Fig.\ref{fig3}. One can easily read that the physical mass of $D_{s0}(2590)$ calculated in the unquenched quark model is 2585 MeV, 88 MeV lower than the original mass 2673 MeV of the bare $c\bar{s}$ state. Compared with Fig.2 of Ref.~\cite{Luo:2021dvj}, we find no similar cusp-like structure as the physical mass $M$ is close to the threshold of $D^*K$, which is only characteristic of $S$-wave couplings and thus called ``$S$-wave threshold effect"~\cite{Rosner:2006vc,Bugg:2008wu}. The main difference in the $\Delta M(M)$ between $S$-wave and $P$-wave two-body coupled channels is owing to the different explicit analytical form of the numerator of the integration, i.e. the transition amplitude $\mathcal{M}(\textbf{p})$ (see Appendix A of Ref.~\cite{Kalashnikova:2005ui}). 

With  the same set of parameters and the mechanism of the $^3P_0$ model, we obtain the decay width into $D^*K$, $\Gamma = 23$ MeV. With $M+i\Gamma/2=2585+i11.5$ MeV, we obtain  $Z = 0.92$, which tells  that the physical $D_{s0}(2590)$ state is dominantly a  $c\bar{s}$ state.

Now we turn to the model independent approach. The $D^*K$ potential of Eq.~(\ref{eq11}) and loop function of Eq.~(\ref{Gs}) contain three unknown parameters $\alpha$, $\beta$, and $\Lambda$. If experimental scattering data existed, as in the cases of $\pi N$~\cite{Aceti:2014ala} and $\pi\pi$~\cite{Aceti:2012dd}, we could have fixed these parameters by fitting to the data. As this is impossible, we could  fit to the experimental mass and width, $M=2591$ and  $\Gamma=89$ MeV. As we only have two data, we cannot determine all the three parameters. Therefore, we choose five different cutoffs $\Lambda=400, 500, 600, 700, 800$ MeV, and try to reproduce the mass and width of $D_{s0}(2590)$ by varying $\alpha$ and $\beta$, and yielding five sets of $\alpha$ and $\beta$ as shown in  Table~\ref{tab:results2}.  In the same table, we also show the obtained pole positions $\sqrt{s_0}=(M,\Gamma/2)$. 

The  $Z$-factor in the model independent approach can be obtained in the following way
\begin{equation}
    -\mathrm{Re}\left[g^2 \left[\frac{d G^{II}(s)}{d s}\right]_{s=s_0}\right]=1-Z,
\end{equation}
where the couplings is calculated as the residue at the pole position 
\begin{equation}
    g^2=\lim_{s\rightarrow s_0}(s-s_0)t^{II}(s).
\end{equation}
The results are given in Table I, One can see that  as the cutoff $\Lambda$ increases from 400 to 800 MeV, the two-body decay width decreases from 80 MeV to 20 MeV and the weight of the $D^*K$ component decreases from 0.58 to 0.05. As we have three unknown parameters but only two data, we cannot tell which cutoff is the optimal one. If we assume that the two-body $D^*K$ decay width almost saturates the $D_{s0}(2590)$ total width, we need a cutoff of 400 MeV. The corresponding weight of the $D^*K$ component is about 60\%. On the other hand, if we believe that the unquenched quark model is correct, i.e., the two-body $D^*K$ decay width is about 20 MeV, we need a cutoff of 800 MeV.  The so obtained weight of the $D^*K$ channel in the physical $D_{s0}(2590)$ state is about 5\%, consistent with 8\% of   the unquenched quark model. Unfortunately, the current experimental data cannot determine the ratio of the two-body decay width with respect to the total decay width~\cite{LHCb:2020gnv}. We hope that future experimental studies can provide such information.
\begin{table}[h!]
    \centering
    \caption{Pole positions, $Z$-factors, couplings, $\alpha$, and $\beta$ obtained with different cutoffs. }\label{tab:results2}
    \begin{tabular}{ccccccc}
    \hline\hline
          $\Lambda$ (MeV) & Pole (MeV) & g & $1-Z$ & $\alpha$~(MeV$^{-2}$)& $\beta$  \\ \hline
                    
          400 & (2591,40) & (0.26,0.11) & 0.58 & $1.2\times10^{-6}$ & 88 \\ 
          500 & (2590, 27) & (0.26,0.07) & 0.25 & $0.85\times10^{-6}$ & 100  \\
          600 & (2590,19) & (0.23,0.04) & 0.05 & $0.7\times10^{-6}$ & 88 \\
          700 & (2590,14) & (0.21,0.03) & 0.02 & $0.78\times10^{-6}$ & 58 \\
          800 & (2590,10) & (0.18,0.02) & 0.05 & $0.59\times10^{-6}$ & 58 \\
                    \hline\hline
    \end{tabular}\\
\end{table}

\section{Summary} 
\label{sec4}
Recently, the LHCb Collaboration reported the discovery of the first radial excited state of $D_{s}$. However, its mass is lower than the quenched  quark model prediction by about 80 MeV, which shows a behavior similar to those of  exotic states such as, $X(3872)$, $D_{s0}^{\ast}(2317)$, and $\Lambda_{c}(2940)$. All of these states have been shown to  couple strongly to the nearby hadronic channels, leading to smaller masses  compared with those of the quenched quark model. In this work, we took into account the $D^{\ast}K$ contribution to the 2$^1S_{0}$ $c\bar{s}$ core to obtain the physical mass of the first excited state of $D_{s}$ in the  unquenched quark model, where the orbital angular momentum of $D^{\ast}$ and $K$  is $P$-wave.  The coupling  of the $c\bar{s}$ core  to $D^{\ast}K$ is estimated by the quark-pair creation model. Our results showed that with only about 10\% of $D^*K$ in the $D_{s0}(2590)$ wave function, the $D^{\ast}K$ contribution could lower the mass obtained  in the quenched quark model by 88 MeV, leading to a mass much closer to the experimental value. The two-body decay width predicted in the same model is about 20 MeV, which only accounts for about one quarter of the total decay width. 

We further constructed a model independent approach to test the unquenched quark model and we found that indeed with a cutoff of about 800 MeV, one can obtain a two-body decay width and a weight of the $D^*K$ component consistent with those of the unquenched quark model. On the other hand, with a cutoff of 400 MeV, one found that the $D^*K$ partial decay width almost saturates the $D_{s0}(2590)$ decay width and the corresponding weight of the $D^*K$ component is about 60\%. Future experimental studies will allow us to fix the unknown parameters of our model and  determine the weight of the $D^*K$ component unambiguously. In particular, a measurement of the partial two-body decay width of $D_{s0}(2590)$ seems to be the key.

\section{Acknowledgments}
We acknowledge useful communications with Liming Zhang and Chen Chen of Tsinghua University. This work is partly supported by the National Natural Science Foundation of China under Grants No.11735003, No.11975041, and No.11961141004, and the fundamental Research Funds
for the Central Universities.
\bibliography{ref}

\appendix
\section{The vacuum quark-pair creation model}
\label{Appendix}

In the following, we provide some details about  the transition amplitude $\mathcal{M}_{c\bar{s}(2^1S_0)\rightarrow D^{\ast}K(\textbf{p})}(\textbf{p})$ in the $^3P_{0}$ model. 
The $c\bar{s}(2^1S_0)$ meson decaying into $D^{\ast}$ and $K$ mesons is allowed in the $^3P_{0}$ model, while the orbital angular momentum between $D^{\ast}$ and $K$ is $\ell$=1. The transition amplitude is the inner product between the initial and final state vectors. The state of the initial meson made up of quark 1 and antiquark 2 has the following form~\cite{PhysRevD.25.1944} 
\begin{equation}
\begin{aligned}
  \Psi^{n\ell sjm_j}_{\textbf{p}_{\rm{tot}}}(q_1 \bar{q}_2)=& \omega^{(12)} \phi^{(12)} \sum_{m_{\ell}m_{s}}C_{\ell s}(jm_j;m_{\ell}m_{s})\chi^{(12)}_{sm_s}\int d^3 p_{1} \int d^3 p_{2} \psi_{n\ell m_\ell}(\textbf{p}_1,\textbf{p}_2)\delta^{(3)}(\textbf{p}_{\rm{tot}}-\textbf{p}_{1}-\textbf{p}_{2})\Psi_{\textbf{p}_{1}}(q_1)\Psi_{\textbf{p}_{2}}(\bar{q}_2)\\
  =& \omega^{(12)} \phi^{(12)} \sum_{m_{\ell}m_{s}}C_{\ell s}(jm_j;m_{\ell}m_{s})\chi^{(12)}_{sm_s}\int d^3 p_{\mathrm{rel}} \psi_{n\ell m_\ell}(\textbf{p}_{\mathrm{rel}})\Psi_{\frac{m_1}{m_1+m_2}\textbf{p}_{\mathrm{tot}}+\textbf{p}_{\mathrm{rel}}}(q_1)\Psi_{\frac{m_2}{m_1+m_2}\textbf{p}_{\mathrm{tot}}-\textbf{p}_{\mathrm{rel}}}(\bar{q}_2),
\end{aligned}
\end{equation}
where $\omega$, $\phi$, and $\chi$ represent  color, flavor, and spin wave functions, respectively. $\psi$ is the wave function in momentum-space. The total and relative 3-momentum of quark 1 and antiquark 2 are defined as the following  
\begin{equation}
    \begin{aligned}
    \textbf{p}_{\mathrm{tot}}&=\textbf{p}_1+\textbf{p}_2, \\
    \textbf{p}_{\mathrm{rel}}&=\frac{m_2\textbf{p}_1-m_1\textbf{p}_2}{m_1+m_2}.
    \end{aligned}
\end{equation}
The final states can be given in the same approach. With these state vectors and the operator $T$ in Eq.~(\ref{eq5}) the helicity amplitude  $\mathcal{M}^{m_{j_{A}}m_{j_{B}}m_{j_{C}}}(\textbf{p})$ for the process $A\rightarrow B+C$ can be written as 
\begin{equation}
\label{hel}
\begin{aligned}
 \mathcal{M}^{m_{j_{A}}m_{j_{B}}m_{j_{C}}}&(\textbf{p})=\gamma \sum_{m_{\ell_{A}} m_{s_{A}} m_{\ell_{B}} m_{s_{B}} m_{\ell_{C}} m_{s_{C}}m}C_{\ell_A s_A}(j_A m_{j_A};m_{\ell_{A}} m_{s_{A}})C_{\ell_B s_B}(j_B m_{j_B};m_{\ell_{B}} m_{s_{B}})\\ \times & C_{\ell_C s_C}(j_C m_{j_C};m_{\ell_{C}} m_{s_{C}})C_{11}(00;m -m) \left(\chi^{(14)}_{s_B m_{s_B}}\chi^{(32)}_{s_C m_{s_C}},\chi^{(12)}_{s_A m_{s_A}}\chi^{(34)}_{1,-m}\right)\\ \times & \left[\left(\phi_{B}^{(14)}\phi_{C}^{(32)},\phi_{A}^{(12)}\phi_{0}^{(34)}\right)\mathcal{I}(\textbf{p},m_1,m_2,m_3)+(-1)^{1+s_A+s_B+s_C}\left(\phi_{B}^{(32)}\phi_{C}^{(14)},\phi_{A}^{(12)}\phi_{0}^{(34)}\right)\mathcal{I}(-\textbf{p},m_2,m_1,m_3)\right],
\end{aligned}
\end{equation}
where the indices 3 and 4 refer to a pair of quark and antiquark created from vacuum, which will form the final states B and C by combining with the quark 1 and antiquark 2 of meson A.   The momentum-space integral $\mathcal{I}(\textbf{p},m_1,m_2,m_3)$  is the overlap of initial and final wave functions
\begin{equation}
    \mathcal{I}(\textbf{p},m_1,m_2,m_3)=\int d^3 k \psi^{\ast}_{n_B \ell_B m_{\ell_B}}\left(\textbf{k}+\frac{m_3}{m_1+m_3}\textbf{p}\right) \psi^{\ast}_{n_C \ell_C m_{\ell_C}}\left(\textbf{k}+\frac{m_3}{m_2+m_3}\textbf{p}\right) \psi_{n_A \ell_A m_{\ell_A}}(\textbf{k}+\textbf{p})\mathcal{Y}_1^m(\textbf{k}).
\end{equation}

In the present work, we choose simple harmonic oscillator~(SHO) wave functions to expand wave functions in momentum-space 
\begin{equation}
    \psi^{\rm{SHO}}_{n \ell m_\ell}(\textbf{p})=R^{\rm{SHO}}_{n\ell}(|\textbf{p}|)Y_{\ell}^{m_\ell}(\hat{p}),
\end{equation}
where the radial wave function is given by
\begin{equation}
\label{sho}
    R^{\rm{SHO}}_{n\ell}(|\textbf{p}|)=\frac{(-1)^n(-i)^\ell}{\beta^{\frac{3}{2}}}\sqrt{\frac{2n}{\Gamma\left(n+\ell+\frac{3}{2}\right)}}\left(\frac{|\textbf{p}|}{\beta}\right)^\ell L_n^{\ell+\frac{1}{2}} \left(\frac{|\textbf{p}|^2}{\beta^2} \right)e^{-\frac{|\textbf{p}|^2}{2\beta^2}}.
\end{equation}
$L_n^{\ell+\frac{1}{2}}$ is the associated Laguerre polynomial and the parameter $\beta_{\rm{eff}}$ is unknown, which can be determined by the requirement that the root-mean-square~(rms) momentum calculated through SHO wave functions should be equal to that of the wave functions calculated using the Godfrey-Isgur relativized potential quark model~\cite{Godfrey:1985xj}.

The color and flavor overlap factors can be readily obtained by the inner product of the corresponding wave functions, of which the details can be found in Appendix of Ref.~\cite{Godfrey:2015dia}.  Spin matrix elements involving the spin of four quarks can be calculated by the angular momentum algebra of Wigner $9j$ symbols
\begin{equation}
\begin{aligned}
 \left(\chi^{(14)}_{s_B m_{s_B}}\chi^{(32)}_{s_C m_{s_C}},\chi^{(12)}_{s_A m_{s_A}}\chi^{(34)}_{1,-m}\right)=&(-1)^{1+s_C}\sqrt{3(2s_A+1)(2s_B+1)(2s_C+1)}\\ &\times \sum_{sm_s}C_{s_B s_C}(s m_s;m_{s_B}m_{s_C})C_{s_A 1}(s m_s;m_{s_A}-m) \left\{ \begin{array}{ccc}
    \frac{1}{2}  & \frac{1}{2} & s_A \\
    \frac{1}{2}  & \frac{1}{2} &  1\\
    s_B  & s_C & s 
 \end{array}\right\}.
\end{aligned}
\end{equation}

Finally, for convenience  we use the Jacob-Wick formula~\cite{Jacob:1959at} to convert the helicity amplitude $\mathcal{M}^{m_{j_{A}}m_{j_{B}}m_{j_{C}}}(\textbf{p})$ into the partial wave amplitude $\mathcal{M}^{\ell s}(|\textbf{p}|)$
\begin{equation}
    \mathcal{M}^{\ell s}(|\textbf{p}|)=\frac{\sqrt{4\pi(2\ell+1)}}{2j_A+1}\sum_{m_{j_B}m_{j_C}}C_{\ell s}(j_A m_{j_{A}}; 0 m_{j_A})C_{j_B j_C}(s m_{j_{A}}; m_{j_B}m_{j_C})\mathcal{M}^{m_{j_{A}}m_{j_{B}}m_{j_{C}}}(|\textbf{p}|\hat{z})|_{m_{j_{A}}=m_{j_{B}}+m_{j_{C}}}.
\end{equation}
Note that we have implicitly assumed that the $z$-axis lies along the direction of the outgoing 3-momentum $\textbf{p}$ of meson B in the final state. For our concrete process $c\bar{s}(2^1S_0)\rightarrow D^{\ast}K(\textbf{p})$, we can relate $\mathcal{M}_{c\bar{s}(2^1S_0)\rightarrow D^{\ast}K(\textbf{p})}(|\textbf{p}|)$ with the partial wave amplitude $\mathcal{M}^{\ell s}(|\textbf{p}|)$ based on the conservation of angular momentum and selection rules for the strong interaction
\begin{equation}
    \mathcal{M}_{c\bar{s}(2^1S_0)\rightarrow D^*K(\textbf{p})}(|\textbf{p}|)=\mathcal{M}^{11}(|\textbf{p}|).
\end{equation}

\end{document}